\documentclass[superscriptaddress,prr, aps,amsmath,amssymb,reprint,showkeys,showpacs
]{revtex4-2}
\usepackage[dvipsnames,x11names]{xcolor}
\usepackage{graphicx, tikz, graphics, epsf, braket}
\usepackage{dcolumn}
\usepackage{bm}
\usepackage[colorlinks=true, citecolor = MidnightBlue, linkcolor = MidnightBlue, urlcolor=MidnightBlue, pdfborder={0 0 0}]{hyperref}
\usepackage{times}
\usepackage{newtxmath}
\usepackage{orcidlink}
\usepackage{nameref}
\usepackage[normalem]{ulem}
\hypersetup{breaklinks=true}
\newcommand{\no}[1]{\langle:\!#1\!:\rangle}
\newcommand{\Gtresmed}{G^{(3/2)}_{c,\,\theta}}
\newcommand{\gtresmed}{g^{(3/2)}_{\theta}}
\newcommand{\gtresmeda}{g^{(3/2)}_{\theta=\arg\langle \hat{a} \rangle}}
\definecolor{changecolor}{RGB}{153,0,255}
\usepackage{fontawesome}

\begin{document}

\preprint{APS/123-QED}

\title{Characterization of Generalized Coherent States through Intensity-Field Correlations}

\author{Ignacio Salinas Valdivieso\, \orcidlink{0000-0002-6912-8606}}
\thanks{These authors contributed equally to this work.}
\affiliation{ANID - Millenium Science Iniciative Program - Millenium Institute for Research in Optics, Chile}

\author{Victor Gondret\,\orcidlink{0009-0005-8468-161X}}
\thanks{These authors contributed equally to this work.}
\affiliation{ANID - Millenium Science Iniciative Program - Millenium Institute for Research in Optics, Chile}
\affiliation{Departamento de F\'isica, Facultad de Ciencias F\'isicas y Matem\'aticas, Universidad de Chile, Santiago, Chile}

\author{Gerd Hartmann S.}
\affiliation{Departamento de F\'isica, Facultad de Ciencias F\'isicas y Matem\'aticas, Universidad de Chile, Santiago, Chile}

\author{Mariano Uria\, \orcidlink{0000-0002-8636-2306}}
\affiliation{Departamento de F\'isica, Facultad de Ciencias F\'isicas y Matem\'aticas, Universidad de Concepci\'on, Concepci\'on, Chile}

\author{Pablo Solano\,\orcidlink{0000-0002-0676-420X}}
\affiliation{Departamento de F\'isica, Facultad de Ciencias F\'isicas y Matem\'aticas, Universidad de Concepci\'on, Concepci\'on, Chile}

\author{Carla Hermann-Avigliano\,\orcidlink{0009-0000-0870-1942}}\homepage{www.amazingquantum.com}\email{carla.hermann@uchile.cl}
\affiliation{ANID - Millenium Science Iniciative Program - Millenium Institute for Research in Optics, Chile}
\affiliation{Departamento de F\'isica, Facultad de Ciencias F\'isicas y Matem\'aticas, Universidad de Chile, Santiago, Chile}

\date{\today}

\begin{abstract}
Non-Gaussian quantum states of light are essential resources for quantum information processing and precision metrology. Among them, generalized coherent states (GCS), which naturally arise from the evolution of a coherent state with a nonlinear medium, exhibit useful quantum features such as Wigner negativity and metrological advantages~\href{http://dx.doi.org/10.1103/PhysRevResearch.5.013165}{[Phys. Rev. Res. \textbf{5}, 013165 (2023)]}. Because these states remain coherent to all orders, their nonclassical character cannot be revealed through standard intensity–intensity correlation measurements. Here, we demonstrate that the intensity–field correlation function alone provides a simple and experimentally accessible witness of nonclassicality. For GCSs, any deviation of this normalized correlation from unity signals nonclassical behavior. We derive analytical results for Kerr-generated states and extend the analysis to statistical mixtures of GCSs. The proposed approach enables real-time, low-complexity detection of quantum signatures in non-Gaussian states, offering a practical tool for experiments across a broad range of nonlinear regimes.

\end{abstract}

\maketitle

\section{Introduction}

Quantum states of light enable measurement sensitivities beyond those achievable with classical resources~\cite{giovannetti.2004.quantumenhanced, pezze.2018.quantum}. 
Practical applications of this quantum advantage include sub-shot noise detection of gravitational waves~\cite{ligo.2011.gravitational} and enhanced precision in microscopy~\cite{he.2023.quantum} using squeezed states of light. 
However, these states are Gaussian with positive Wigner functions~\cite{walschaers.2021.nonGaussian}, whereas other non-Gaussian quantum states can exhibit negative values in their Wigner functions. Such negativity can enhance phase estimation in metrology~\cite{lupu-gladstein.2022.negative}, and non-Gaussian quantum states have been shown to outperform their Gaussian counterparts in many configurations~\cite{oh.2019.optimal, woodworth.2020.transmission, marolleau.2024.subshotnoise, park2025quantumphaseestimationgaussian, woodworth.2022.transmission}.
In addition, in quantum information, Wigner negativity is a key resource for continuous-variable quantum computation~\cite{mari.2012.positive}, making the generation and detection of such states a central challenge in modern quantum optics and quantum technology~\cite{walschaers.2021.nonGaussian}.

Non-Gaussian states can be generated either probabilistically, for example, through photon subtraction, or deterministically via nonlinear interactions~\cite{lvovsky.2020.production}. 
In cavity quantum electrodynamics, coupling a cavity mode to a single atom allows the deterministic production of large Fock states~\cite{sayrin.2011.realtime, uria.2020.deterministic}. 
In the optical regime, interactions of light with a single atom~\cite{hacker.2019.deterministic} or a cold cloud of Rydberg atoms~\cite{magro.2023.deterministic} have enabled deterministic generation of free-propagating Wigner-negative states. 
Other nonlinearities, such as Kerr-type interactions, can be used to produce non-Gaussian states~\cite{puri.2017.engineering, strandberg.2021.wigner, shalibo.2013.direct} including Schr\"odinger cat states~\cite{yurke.1986.generating, kirchmair.2013.observation}. 
More generally, the evolution of a coherent state under a Hamiltonian $\hat{H}_{\mathrm{int}} \propto \hat{n}^{\varepsilon}$, nonlinear in the photon-number operator $\hat{n}$, yields a non-Gaussian state whenever the nonlinear parameter $\varepsilon$ is different from both 0 and 1~\cite{lloyd.1999.quantum}. 
These states are well described by the family of generalized coherent states (GCSs)~\cite{uria.2023.emergence, Stoler_1971}. 
They are termed ``coherent'' because they preserve the same particle-number statistics as standard coherent states, remaining coherent to all orders in the sense of Glauber~\cite{glauber.2006.Quantum}. 
Remarkably, despite not being minimum-uncertainty states, GCSs can exhibit strong nonclassical features, including Wigner negativity and significant metrological advantages~\cite{uria.2023.emergence}.

Quantifying the nonclassicality of a quantum state can be done using its Wigner function negativity~\cite{kenfack.2004.negativity}, entanglement potential~\cite{asboth.2005.computable}, or stellar rank~\cite{chabaud.2020.stellar, chabaud.2021.certification}. 
However, these measures typically require many measurements, detailed data analysis, and can be affected by non-perfect efficiencies, as for the negativity of the Wigner function~\cite{lvovsky.2002.nonclassical, yurke.1986.generating}. 
Instead, only a few correlation functions  can be sufficient to reveal nonclassicality~\cite{carmichael.2000.giant, carmichael.2004.intensityfield, shchukin.2005.nonclassicality, shchukin.2005.nonclassical, sperling.2013.correlation, rivas.2009.nonclassicality, laiho.2022.measuring}. 
Because GCSs are coherent to all order, their nonclassicality cannot be revealed using only intensity-intensity correlation functions, i.e. correlation functions which involve the same number of creation and annihilation operators $\hat{a}^\dagger$ and $\hat{a}$.
Their nonclassicality is instead encoded in their correlation functions involving different numbers of $\hat{a}^\dagger$ and $\hat{a}$. This stems from the fact that a quantum state at a given time is completely characterized by the set of all its equal-time correlation functions of the field operators~\cite{zache.2020.extracting}.

In this work, we explore the use of a low-order correlation function---specifically, the intensity–field correlation function---to probe the nonclassical nature of GCSs. 
We show that, for GCSs, any deviation from unity of the normalized intensity-field correlation function $\gtresmed$  [see Eq.~\eqref{eq:def_g32}] directly signals the nonclassicality of the state.
The main advantage of this observable lies in its relatively straightforward experimental implementation, illustrated schematically in Fig.~\ref{fig1}. This contrast with the large number of detectors or time-multiplexing schemes typically required to measure higher-order correlation functions~\cite{avenhaus.2010.accessing}. 
Furthermore, the evaluation of this function does not require long post-acquisition data analysis allowing for real time measurement and is thus a practical tool for experimentalists.

The paper is organized as follows. In section~\ref{secII}, we define the intensity–field correlation function and relate it to nonclassical signatures. Section~\ref{secIII} applies it to GCSs and focuses on a simple illustrative example—the Kerr state—where the expression for $\gtresmed$ takes a simple form. Section~\ref{secIV} studies the intensity-field correlation function for different nonlinear parameters. Additionally, it shows the robustness of the method with non-perfect efficiency or decoherence effects and discusses its extension to the large field intensity limit. Finally, we conclude in section~\ref{secV}.

\section{Intensity-field correlation function as a probe of nonclassicality}\label{secII}

\begin{figure}
\centering
\includegraphics[width=3.375in]{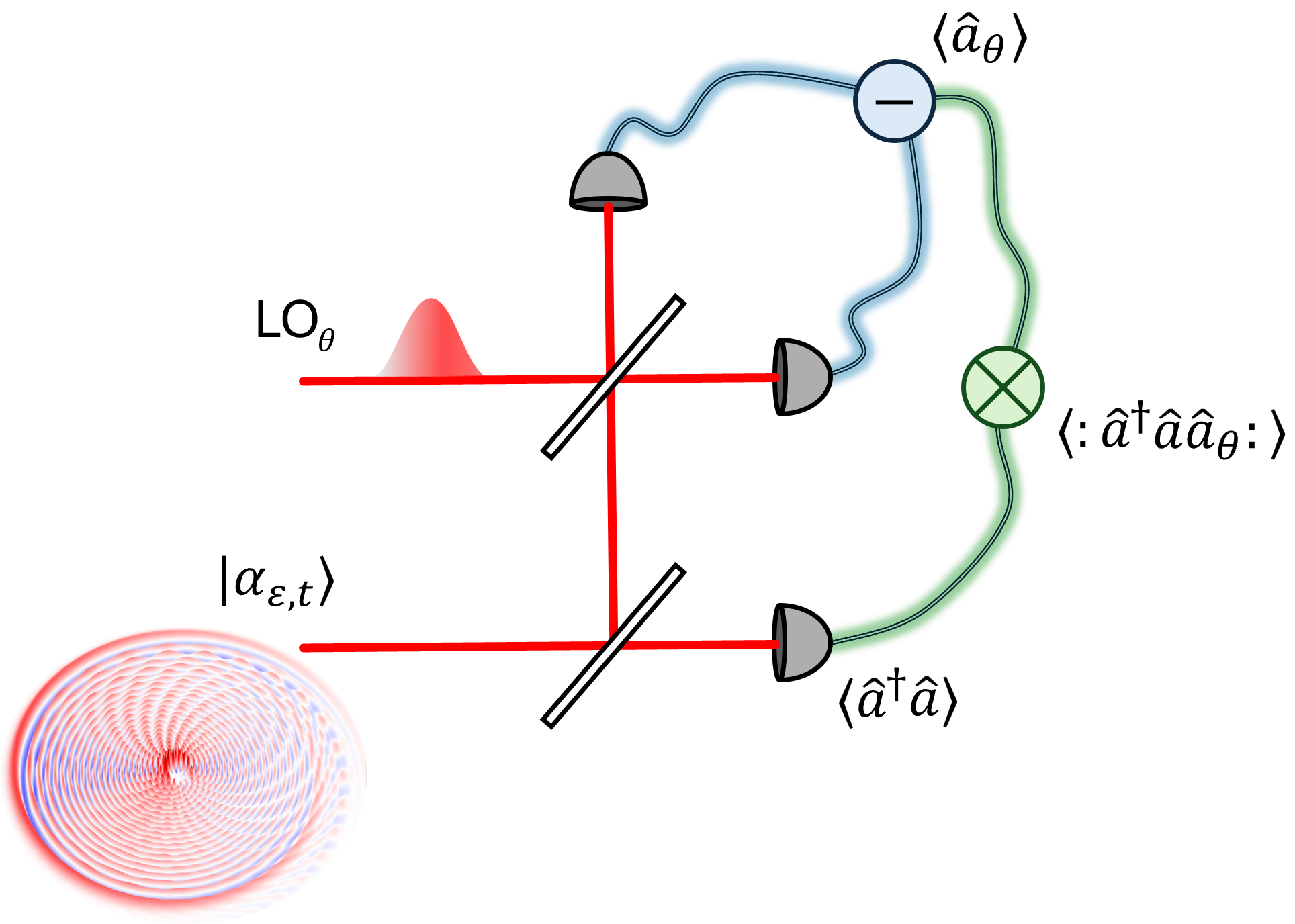}
\caption{
Optical setup to measure the intensity-field correlation function. 
A GCS state $\ket{\alpha_{\varepsilon,t}}$ is split on a beam-splitter. Part of the state is sent to an intensity detector and the other part is mixed with a local oscillator (LO$_\theta$) to measure the quadrature along $\theta$.}
\label{fig1}
\end{figure}

The intensity-field correlation function was first introduced by Carmichael and collaborators~\cite{carmichael.2000.giant, carmichael.2004.intensityfield} in the context of conditional homodyne detection. 
This observable arises naturally in a measurement setup based on a modified Hanbury Brown and Twiss interferometer~\cite{hanbury.1956.correlation}, where one of the detectors is replaced by a balanced homodyne detector that measures the quadrature along $\theta$~\cite{OROZCO1998}, see Fig.~\ref{fig1}. 
Although the time dependence of this function can reveal nonclassicality by introducing a delay between the intensity and homodyne detectors~\cite{denisov.2002.timeasymmetric, gutierrez.2017.fluctuations, castro.2024.quantum, santo-2021.phase}, we focus here on the normalized equal-time intensity-field correlation function
\begin{equation}
 g^{(3/2)}_\theta =\frac{\no{ \hat{n}\hat{a}_\theta}}{\langle \hat{n}\rangle\langle\hat{a}_\theta\rangle}  ,
    \label{eq:def_g32}
\end{equation}
where $\hat{n}=\hat{a}^\dagger\hat{a}$ is the number operator and $\hat{a}_\theta=(\hat{a}e^{-i\theta}+\hat{a}^\dagger e^{i\theta})/2$ is the quadrature operator along the direction $\theta$, the phase of the local oscillator. The colon symbols indicate that the operators between them must be normal-ordered.
Note that for definiteness of the $g^{(3/2)}_\theta $ function, the quadrature probed should not be orthogonal to the state displacement so that $\braket{\hat{a}_\theta}\neq 0$.

Reference~\cite{carmichael.2000.giant} showed that if the phase of the local oscillator matches that of the signal,  ($\arg\langle \hat{a}\rangle=\theta$) and if the state is Gaussian, then, the value of $\gtresmeda$ is bounded. 
In this approach, the derivation of the nonclassical bound relies on the positivity of (normal ordered) variances operators for any classical state. 
Fluorescence signal from atoms in cavity quantum electrodynamics~\cite{foster.2000.quantum} and a single ions~\cite{gerber.2009.intensityfield} then reported an experimental observation of the violation of this classical bound, thereby demonstrating the quantumness of the emitted light.
\newline However, this approach possesses severe drawbacks as it requires to measure the phase of the state prior to the measurement of $\gtresmeda$ and the nonclassicality bound only applies to Gaussian state, limiting its applicability to GCSs.
An alternative path relies on considering other correlators than Eq.~\eqref{eq:def_g32}, following the approach of Shchukin and Vogel~\cite{shchukin.2005.nonclassical,shchukin.2005.nonclassicality}.
In their work, they derive a hierarchy of inequalities involving moments of creation and annihilation operators, whose violation is a sufficient condition for nonclassicality. 
Here, the notion of classicality is based on a well-defined positive Glauber-Sudarshan $P$ representation~\cite{glauber.1963.coherent, sudarshan.1963.equivalence}. 
In particular, 
they show that any classical state must satisfy $ D_\theta^{(3)}\geq 0$, where~\cite{shchukin.2005.nonclassicality}
\begin{equation}
    D_\theta^{(3)}=\begin{vmatrix}
        1 & \langle\hat{a}_\theta\rangle & \langle\hat{n}\rangle\\
        \langle\hat{a}_\theta\rangle &\no{\hat{a}^2_\theta} & \no{\hat{a}_\theta\hat{n}} \\
        \langle\hat{n}\rangle& \no{\hat{a}_\theta\hat{n}} & \no{\hat{n}^2}
    \end{vmatrix}.\label{eq:def_D3}
\end{equation}
For our purpose, it is enlightening to rewrite this determinant as
\begin{equation}
\begin{split}
 D_\theta^{(3)}= & \left(\no{\hat{a}_\theta^2} - \braket{\hat{a}_\theta}^2\right)\left( \no{\hat{n}^2} - \braket{\hat{n}}^2\right) \\
 & -\braket{\hat{a}_\theta}^2\braket{\hat{n}}^2 \left(\gtresmed-1 \right)^2.
\end{split}
\label{eq:expended_D3}
\end{equation}
On the right-hand side of Eq.~\eqref{eq:expended_D3}, the sign of the first line can vary, but the second line is either null or negative. Nonclassicality witnessed by  $D_\theta^{(3)}<0$ can thus occur if the sign of the two terms of the first line are opposite or if the absolute value of the second line exceeds the first one. \newline
Negativity of $D_\theta^{(3)}$ occurs for squeezed vacuum states when the quadrature probed is aligned with the anti-squeezed direction. In this scenario, the first term is negative ($\no{\hat{a}^2_\theta}<0$) while the second term is positive (superbunching). The determinant $D_\theta^{(3)}$ is also negative for Fock states. In this case, the first term of the first line is positive, while the second is negative (anti-bunching).

\section{Application to Generalized Coherent States}\label{secIII}
\subsection{General case}\label{secIII.A}
GCSs are coherent to all orders~\cite{Stoler_1971}, therefore ${\no{\hat{n}^2} - \braket{\hat{n}}^2}$ vanishes and so does the first line of Eq.~\eqref{eq:expended_D3}. 
Hence, any deviation of $\gtresmed$ from unity leads to a negative determinant $D_\theta^{(3)}<0$, thereby implying the nonclassicality of state. 
The validity of the nonclassical witness $D_\theta^{(3)}$ is not conditioned to the phase matching between the signal and the local oscillator, as required in the original witness of Carmichael \textit{et al.}~\cite{carmichael.2000.giant}, although it requires a stable phase $\theta$ during the measurement.

For coherent states $\ket{\alpha}$, it is straightforward to show that $g_\theta^{(3/2)}=1$, no matter the probed quadrature~$\theta$. For these states, $D_\theta^{(3)}$ vanishes thereby confirming coherent states are the limit between the quantum and classical worlds. 
We introduce the following GCS family~\cite{uria.2023.emergence}
\begin{equation}
\ket{\alpha_{\varepsilon, t}} = e^{-\frac{|\alpha|^{2}}{2}} 
\sum_{n=0}^{\infty} 
\frac{\alpha^{n}}{\sqrt{n!}} 
e^{-i t n^{\varepsilon}} 
\ket{n}_\text{\scriptsize F},
\label{eq:psiGCS}
\end{equation}
where $\ket{n}_\text{\scriptsize F}$ denotes the Fock state basis and the other parameters are introduced hereafter.
GCS family \eqref{eq:psiGCS} was shown by Ref.~\cite{uria.2023.emergence} to be the natural basis to describe the evolution of an initial coherent state of amplitude $\alpha$ under a nonlinear Hamiltonian of the form $\hat H_\mathrm{int} = \hbar\kappa \hat n ^\varepsilon$. Here, $\varepsilon$ refers to the nonlinear medium power and $t=\kappa\Delta t$ is a dimensionless quantity, the product of the interaction strength $\kappa$ and the interaction duration of the light with the nonlinear medium $\Delta t$.
When omitting subscripts in $\ket{\alpha_{\varepsilon, t}}$, we will implicitly assume they are null so that $\ket{\alpha}$ refers to a coherent state. Note that the most general expression for GCSs has arbitrary phase parameters multiplying each Fock state~\cite{Stoler_1971}, so Eq.~\eqref{eq:psiGCS} only refers to a subset of all pure GCEs.

We use Eq.~\eqref{eq:psiGCS} to compute the correlators of Eq.~\eqref{eq:def_g32} where each term is given by~\cite{uria.2023.emergence}
\begin{equation}
     \begin{split}
        \langle\hat{a}_\theta\rangle =& \, e^{-|\alpha|^2}\sum_{n=0}^\infty\frac{|\alpha|^{2n+1}}{n!}\cos{(\varphi -\theta  -z_{n,1} (t)+z_{n,0}(t))},\\
    \no{\hat{n}\hat{a}_\theta} =& \,  e^{-|\alpha|^2}\sum_{n=0}^\infty\frac{|\alpha|^{2n+3}}{n!}\cos(\varphi -\theta  - z_{n,2}(t)+z_{n,1}(t)),\\
    \end{split}\label{eq:a_g32_GCS}
\end{equation}
where $z_{n,j}(t)=t(n+j)^\varepsilon$ and $\varphi=\arg(\alpha)$ is the phase of the initial state and $\langle\hat{n}\rangle=|\alpha|^2$.
A further simplification of the summation is in general not possible. However, for specific values of the nonlinear parameter, a simpler expression can be obtained. \newline

\subsection{The case of Kerr states}
When $\varepsilon=2$, i.e. when the interaction Hamiltonian is quadratic in the number operator (Kerr Hamiltonian), the GCSs reduce to Kerr states. In this specific case, Eq.~\eqref{eq:a_g32_GCS} is given by
\begin{equation}
    \begin{split}
        \langle\hat{a}_\theta\rangle =& \, |\alpha|e^{-|\alpha|^2(1-\cos(2t))}\cos\left(\varphi - \theta -t -|\alpha|^2\sin2t\right),\\
   \no{\hat{n}\hat{a}_\theta} =&  \, |\alpha|^3e^{-|\alpha|^2(1-\cos(2t))}\cos\left(\varphi - \theta -3t -|\alpha|^2\sin2t\right),\\
    \end{split}\label{eq:a_g32_Kerr}
\end{equation}
and the normalized intensity-field correlation function is simply given by
\begin{equation}
\gtresmed=\cos(2t)+\sin(2t)\tan\left(\varphi - \theta -t -|\alpha|^2\sin(2t)\right).
\label{eq:gtresmed_kerr}\end{equation}
When $t$ is an integer multiple of $\pi$, this function equals 1, no matter the value of $\theta$. This is because Kerr states periodically return to coherent states for which $\gtresmed=1$~\cite{kirchmair.2013.observation}. 
However, as long as $t$ is not an integer multiple of $\pi$, a strong violation of $g^{(3/2)}_\theta=1$ can occur, which evidences the nonclassicality of the state. 
This nonclassicality can be further evidenced by the negativity of the state’s Wigner function~\cite{kirchmair.2013.observation} and by its quantum Fisher information~\cite{uria.2023.emergence}.

The phase inside the tangent function  corresponds to the phase difference between the local oscillator phase $\theta$ and the phase of the state given by $\arg \langle \hat{a} \rangle = \varphi-t-|\alpha|^2\sin 2t$. 
When this phase difference is $\pm\pi/2$, the probed quadrature is perpendicular to the displacement of the state and thus $\langle \hat{a}_\theta \rangle=0$, which is responsible for the divergence of~$g^{(3/2)}_\theta$. 
We note however that nonclassicality, based on the determinant of Eq.~\eqref{eq:expended_D3}, is guaranteed whenever 
$\no{\hat{n}\hat{a}_\theta}$ differs from $\langle \hat{n}\rangle\langle\hat{a}_\theta\rangle$.
Therefore, if the quadrature probed is such that  $\langle \hat{a}_\theta \rangle=0$, a non-zero value of the three field correlator $\no{\hat{n}\hat{a}_\theta}$ is a sufficient condition for nonclassicality.

\section{Analysis}\label{secIV}
\subsection{\boldmath $\gtresmed$ for different nonlinear parameters and interaction times}

\begin{figure}[htbp]
    \centering
    \includegraphics[width=3.375in]{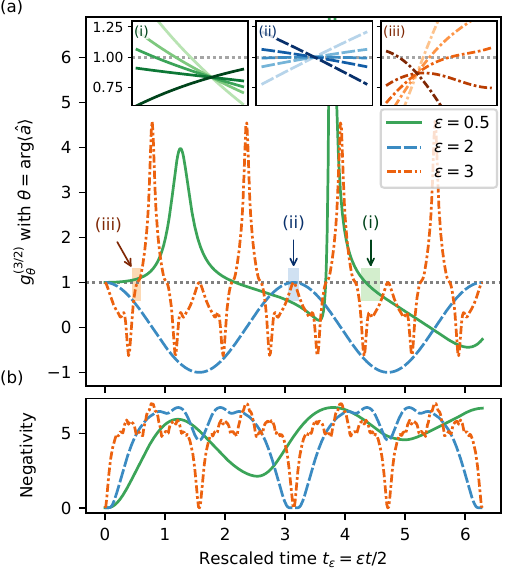}
    \caption{Normalized intensity-field correlation function (a) and Wigner negativity (b) as a function of the rescaled time $t_\varepsilon = \varepsilon t /2$ for $\varepsilon=0.5$ (solid green), $\varepsilon=2$ (dashed blue) and $\varepsilon=3$ (dashed-dotted orange).  The dashed gray line corresponds to the classical limit. 
    The insets show the  time evolution of $\gtresmed$ for different local oscillator phases $\theta =\arg\langle\hat{a}\rangle+\delta\phi$, near the time point where $\gtresmeda$ crosses the classical limit 1. The color intensity encodes the supplemental phase $\delta\phi=-\pi/3,-\pi/6,0, \pi/6,\pi/3$ from lighter to darker, respectively. Here the mean particle number is $\langle\hat{n}\rangle=1$. The negativity is computed integrating the negative values of the Wigner function using QuTIP~\cite{lambert.2026.qutip}.}
    \label{fig2}
\end{figure}
Without loss of generality, and for the purpose of simplifying the visualization and analysis, we choose the phase of the probed quadrature to match the one of the state, i.e. $\theta = \arg \langle \hat{a} \rangle$. 
We show in Fig.~\ref{fig2}(a) the evolution of $\gtresmeda$ as a function of a rescaled effective time $t_\varepsilon = \varepsilon t /2$ (for visualization purposes) for three values of the nonlinear parameter~$\varepsilon$.
The gray dashed line, which could corresponds to a nonlinear parameter of $\varepsilon=1$ or $0$, shows the value of $\gtresmed=1$ for coherent states i.e. representing the classical limit.
Here, we observe that for most of the time, the value of $\gtresmeda$ of GCSs differs from 1, revealing the state's nonclassicality.
This nonclassicality is corroborated by the negativity of the states, plotted in Fig.~\ref{fig2}(b). Here, the negativity  is defined as the integral of the negative part of the state's Wigner function: a nonzero value of the negativity is different indicator of the nonclassical character of the state~\cite{kenfack.2004.negativity}. Note that the comparison between the $\gtresmed\neq 1$ criterion and Wigner negativity is provided only for illustration purposes, as no formal connection between the two has been established.

For Kerr states ($\varepsilon=2$, dashed blue line), the expression of $\gtresmeda$ is further simplified to $\cos(2t)$, as can be shown from Eq.~\eqref{eq:gtresmed_kerr}.
Nonclassicality of Kerr states fades away after a time which is an multiple integer of~$\pi$, indicated by arrow (ii) in Fig.~\ref{fig2}. This is because these states periodically return to coherent states~\cite{yurke.1986.generating, kirchmair.2013.observation}.
GCSs states with $\varepsilon=3$ (dashed-dotted orange) also periodically return to coherent states albeit with a period shorter by a factor 3.
However, we observe that both for $\varepsilon=3$ and $\varepsilon=0.5$ (solid green), the $\gtresmeda$ curve crosses the unity line  at times for which the Wigner negativity does not vanish, see arrows (i) and (iii). At these specific times~$t^\star$, the $\gtresmeda$ function fails to capture the nonclassicality of the state.
In the insets of Fig.~\ref{fig2}(a), we show the time evolution of $\gtresmed$ in the vicinity of~$t^\star$ varying the local oscillator phase $\theta =\arg\langle\hat{a}\rangle+\delta\phi$, with $\delta\phi$ ranging from $-\pi/3$ (lighter) to $+\pi/3$ (darker).
We see that in (i) and (iii), the $\gtresmed$ curves do not collapse on~1 at~$t^\star$. This means that in these cases, the quantumness of the state can still be revealed measuring $\gtresmed$ with a different local oscillator phase~$\theta$. This is not the case in (ii), where all $\gtresmed $ curves collapse on 1 at $t=\pi$ because the state is truly classical i.e., it returned to a coherent state~\cite{kirchmair.2013.observation,yurke.1988.dynamic}.

\subsection{Intense light regime}

For Kerr-like states, Eq.~\eqref{eq:gtresmed_kerr} also reveals an interesting property: the value of $\gtresmeda$ is independent of the mean population.
This behavior contrasts, for example, with that of Fock states $\ket{n}$, whose normalized two-body correlation function approaches the classical limit as the population increases, $g^{(2)} = 1 - 1/n \to 1$ for large $n$, even though their metrological advantage grows with increasing population~\cite{deng.2024.quantumenhanced}. 
We note, however, that as the mean population $|\alpha|^2$ increases, both $\langle \hat{a}_\theta \rangle$ and $\no{\hat{n}\hat{a}_\theta}$ exponentially approach zero, as seen from Eq.~\eqref{eq:a_g32_Kerr}. 
Consequently, although the value of $\gtresmed$ itself is not affected by increasing population, accurately measuring both the numerator and denominator becomes more challenging due to their reduced amplitude.

\begin{figure}
    \centering
    \includegraphics[width=3.375in]{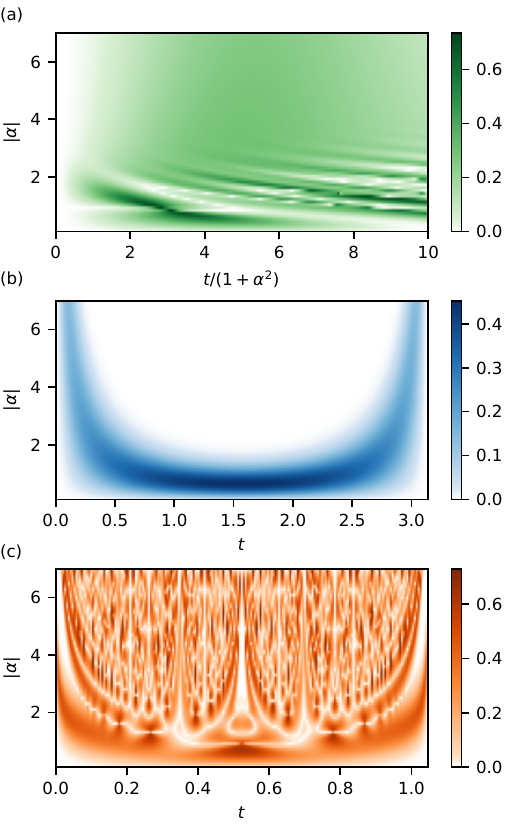}
    \caption{ Evolution of $|\Gtresmed|/|\alpha|^{3/2}$ as a function of time $t$ and the initial displacement  $|\alpha|$ for $\varepsilon=0.5$ (a), $2$ (b) and $3$ (c). A positive value witnesses nonclassicality of the CGS and the intensity of the color encode the value of $|\Gtresmed|/|\alpha|^{3/2}$. The colors of the heatmaps match the ones of Fig.~\ref{fig1}.}
    \label{fig3bis}
\end{figure}

To investigate how this nonclassical witness is affected by the intensity, we introduce the following (non-normalized) connected correlation function
\begin{equation}
    \Gtresmed=\no{\hat{n}\hat{a}_\theta}-\langle \hat{n}\rangle\langle\hat{a}_\theta\rangle. \label{eq:def_Gc32}
\end{equation}
Referring to $\Gtresmed$ as a connected correlation (or cumulant) is somewhat abusive, as a strictly connected three-operator correlation would require subtracting all contributions built from lower-order moments~\cite{kubo.1962.generalized}. Nevertheless, it removes the trivial contributions associated with $\hat{n}$ and $\hat{a}_\theta$ taken alone, in the same spirit as connected particle-number correlations commonly used in cold-atom experiments~\cite{chalopin.2025.connected}. This definition is motivated by  Eq.~\eqref{eq:expended_D3} which shows that any nonzero value of $\Gtresmed$ guarantees nonclassicality of GCSs.  
We show in Fig.~\ref{fig3bis} the value of $|\Gtresmed|/|\alpha|^{3/2}$ as a function of $t$ and $\alpha$ for $\theta=\arg\braket{\hat{a}}$. The darker, the greater the value of $|\Gtresmed|/|\alpha|^{3/2}$ and thus the easier the experimental detection of nonclassicality. \newline
For $\varepsilon=0.5$ (panel a), the  time evolution is not periodic. In the figure,  time is rescaled with the particle number because the nonclassical features take longer to develop when $|\alpha|$ increase, thereby limiting the experimental generation of highly negative quantum states~\cite{uria.2020.deterministic}. 
In the time-range shown here, the intensity-field correlation function accurately captures the nonclassicality of the state even when $|\alpha|$ increases.\newline
For $\varepsilon=2$ and $3$ (panels b and c), the time evolution is periodic and is shown from 0 to $\pi$ and $\pi/3$, respectively.  
For Kerr states, we observe that, when $|\alpha|$ increases, the value of $\Gtresmed$ approach zero for a large time interval, except when $\cos 2t\lesssim 1$ [see the large white region in Fig.~\ref{fig3bis}(b)].
This is in contrast with the expected visibility of the normalized intensity-field correlation function, which is the greatest for $t\sim\pi/2$, see Fig.~\ref{fig2}.
Here the small value of $|\Gtresmed|$ confirms that the visibility of nonclassicality vanishes.
In this case, higher order correlation functions would be better suited to detect the nonclassical behavior of Kerr-like CGS. 
This can be seen from Eq.~\eqref{eq:a_g32_Kerr}, where a higher order correlation function would increase the exponent of $|\alpha|^3$. \newline
For $\varepsilon=3$, we observe the appearance of oscillations as the particle number $|\alpha|^2$ increases. A similar pattern appears when plotting $|\gtresmeda-1|$.
This oscillation is due to the rapidly oscillating phase in Eq.~\eqref{eq:a_g32_GCS}. 
However, Fig.~\ref{fig3bis}(c) shows that the contrast of the oscillation does not decrease with $|\alpha|$.

\subsection{Statistical mixture of GCSs}
The GCSs considered in Eq.~\eqref{eq:psiGCS} are pure states that form a subset of the complete family of GCSs~\cite{Stoler_1971}.
On the other hand, due to the matter–field interaction itself or to losses, for example, the produced state can also be mixed, while still retaining full Glauber coherence~\cite{glauber.1963.coherent}.
This can be illustrated by considering the interaction of light with a single atom in a cavity: during the evolution, the two modes become entangled and later periodically disentangle. 
Tracing over the atomic subsystem in such an entangled state therefore yields a mixed state for the light mode~\cite{hacker.2019.deterministic, uria.2020.deterministic}.\newline
Ref.~\cite{uria.2023.emergence} showed that the evolution of a coherent state $\ket{\alpha}$ under a nonlinear Hamiltonian can also yield a statistical mixture of GCSs all of which share the same amplitude $|\alpha|$. 
Although these states are not pure, they can still exhibit both Wigner-function negativity and metrological advantage~\cite{uria.2023.emergence}. 
A statistical mixture $\hat{\rho}=\sum_i p_i\hat{\rho}_{i}$ of pure GCSs  $\hat{\rho}_i=\ket{\alpha_{\varepsilon_i, t_i}}\bra{\alpha_{\varepsilon_i, t_i}}$ sharing the same amplitude $\alpha$ preserves the underlying Poissonian distribution and therefore remains a GCS. As such, the nonclassicality condition $\Gtresmed \neq 0$ continues to apply.
Furthermore, because 
$\langle \hat{a}^\dagger\hat{a}\rangle = |\alpha|^2$ for each $\hat{\rho}_i$, 
the connected correlation function  $\Gtresmed$ is linear in the mixture i.e. $\Gtresmed(\hat{\rho})=\sum_ip_i\Gtresmed(\hat{\rho_i})$. 
Therefore, it is possible that even though each $\Gtresmed(\hat{\rho_i})$ is nonzero, its weighted sum could be zero so that the quantumness of $\hat{\rho}$ would not be revealed by the measurement of $\Gtresmed$. \newline
A statistical mixture of GCSs that do not share the same amplitude---i.e., the same photon statistics---is not in general a GCS. For such states, the measurement of $\Gtresmed$ alone cannot assess the nonclassicality of the state, and one must instead measure all the terms entering the determinant in Eq.~\eqref{eq:def_D3}. Thus, the  $\Gtresmed\neq 0$ criterion is not robust against Gaussian additive noise channels, as these introduce random displacements in phase space and thus disturb the Glauber coherence of the state. Instead, other experimentally relevant noises such a pure losses modify the particle statistics while preserving the Glauber coherence of the state. The following section illustrates the effect of a pure loss channel with a simple example, where we consider the damping of an initially pure Kerr state.

\subsection{Photon-loss model}\label{sec:losses}

To model the effect of losses on GCSs, we use an amplitude-damping channel described by Kraus operators~\cite{Liu.2004.kraus}. In this model, decoherence is characterized by a single parameter, $\eta$, which can be expressed as $\eta = e^{-\gamma \tau}$, where $\gamma$ is the cavity decoherence rate in which the state is stored~\cite{kirchmair.2013.observation}, and $\tau$ is the evolution time. This description is equivalent to modeling the system as passing through a beam splitter of transmissivity $\sqrt{\eta}$, mixing the state with the vacuum and tracing out the ancillary mode~\cite{escher.2011.quantum,Chuang.1997}.
In appendix \ref{app:A}, we show that such a decoherence channel affects the density matrix of an initial GCS $\hat{\rho}=\ket{\alpha_{\varepsilon, t}}\bra{\alpha_{\varepsilon, t}}$ as  
\begin{equation}
     \hat{\rho}_\eta=\sum_{k=0}^{\infty}p_{\eta,k} \ket{\sqrt{\eta}\alpha_{\varepsilon, t,k}} \bra{\sqrt{\eta}\alpha_{\varepsilon, t,k}}\label{eq:decoherence_GCS}
\end{equation}
where $p_{\eta,k}$ is given in the appendix and physically corresponds to a $k$-photon loss probability and where we have generalized the definition \eqref{eq:psiGCS} of GCSs so that
\begin{equation}
    \ket{\alpha_{\varepsilon, t,k}}= e^{-\frac{|\alpha|^{2}}{2}} 
\sum_{n=0}^{\infty} 
\frac{\alpha^{n}}{\sqrt{n!}} 
e^{-i t (n+k)^{\varepsilon}} 
\ket{n}_\text{\scriptsize F}.\label{eq:anootherGCS}
\end{equation}
The states in \eqref{eq:anootherGCS} are pure and also GCSs because their probability distribution remains poissonian. When $k=0$, we recover the subset introduced in Eq.~\eqref{eq:psiGCS}.\newline
Under a loss channel, Eq.~\eqref{eq:decoherence_GCS} shows that an initial pure GCS transforms into a statistical mixture of GCSs whose amplitude have been damped by a factor $\sqrt{\eta}$. 
However, the statistical mixture Eq.~\eqref{eq:decoherence_GCS} involves GCSs sharing the same amplitude $\sqrt{\eta}\alpha$ which implies that the state remain coherent throughout its evolution.
In the case where the initial state at $\tau=0$ is a coherent state, Eq.~\eqref{eq:decoherence_GCS} simplifies and reduces to a pure coherent state albeit with a damped amplitude.
In the case where the initial state is not coherent, the state becomes mixed and loses its nonclassical features while evolving toward the vacuum state.

\begin{figure}
    \centering
    \includegraphics[width=3.375in]{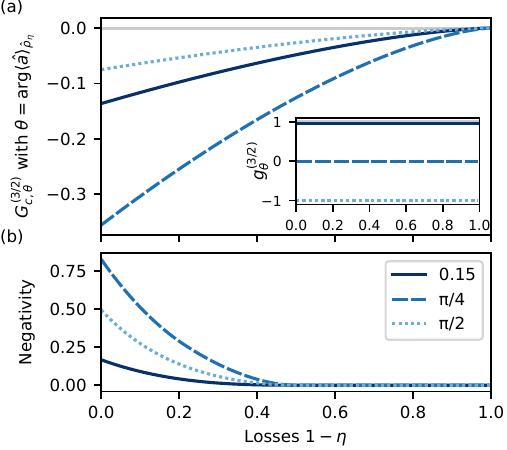}
    \caption{Connected intensity-field correlation function (a) and Wigner negativity (b) as a function of losses $1-\eta$ for different Kerr state parametrized by $t=0.15,\, \pi/4,$ and $ \pi/2$, respectively plotted with solid, dashed and dotted lines. The inset shows the normalized intensity field correlation function, which is constant up to $\eta=0$, for which the value is 1. The negativity is computed integrating the negative values of the Wigner function using Julia's QuantumToolbox~\cite{QuantumToolbox.jl2025}.}
    \label{fig4}
\end{figure}
To illustrate the effect of this decoherence channel in a practical situation,
we consider an initial Kerr state $\ket{\alpha_{2,t}}$, such as the one prepared in Ref.~\cite{kirchmair.2013.observation}.
Figure~\ref{fig4}(a) shows the evolution of $\Gtresmed$ as a function of the loss parameter $1-\gamma$, which ranges from 0 (initial state) to 1 (vacuum state). For comparison, Fig.~\ref{fig4}(b) shows the corresponding evolution of the state’s negativity.
Both witnesses converge monotonically toward 0, the expected value for the vacuum state.
However, we observe that a large value of $|\Gtresmed|$ does not imply a large negativity.
This is simply a consequence of the fact that the determinant~\eqref{eq:def_D3} does not quantify the nonclassicality of the state~\cite{shchukin.2005.nonclassical}.
We also observe that the intensity-field correlation function $\Gtresmed$ continues to capture the state’s nonclassicality for $\eta < 0.5$, whereas the Wigner negativity fails to do so due to its strong sensitivity to losses~\cite{lvovsky.2001.quantum, lvovsky.2002.nonclassical}.
The ability of the intensity-field correlation function to capture nonclassicality is further illustrated by examining the evolution of the normalized correlation function $\gtresmeda$, shown in the inset of Fig.~\ref{fig4}(a).
We observe that $\gtresmeda$ is constant, as expected for a normalized correlation function under a pure loss channel. Its value reaches unity only for $\eta = 0$, when the state becomes the vacuum.
Because $\Gtresmed = (\gtresmed - 1)\langle \hat{a}^\dagger \hat{a} \rangle \langle \hat{a}_\theta \rangle$, the fact that $\gtresmeda$ remains constant shows that the nonclassicality captured by the intensity-field correlation function is not altered by losses once normalized by the particle number and $\langle \hat{a}_\theta \rangle$.
This result further illustrates that negativity and $\gtresmed$ capture different types of nonclassical features.

\section{Conclusion}\label{secV}

In this work, we have investigated the potential of the intensity–field correlation function as a tool for detecting quantum signatures in non-Gaussian states, focusing on the family of generalized coherent states. Because these states are coherent to all order in the sense of Glauber, they cannot be distinguished from traditional coherent states with traditional correlation functions.
We have shown that the nonclassical character of GCSs can be readily identified by measuring the intensity–field correlation function over a wide range of nonlinearities.
This witness is experimentally simple to implement and the straightforward nature of the data analysis, enabling measurements to be performed almost in real time, is a practical tool for experimentalists.

\section*{Acknowledgments} 
This work was supported in part by FONDECYT REGULAR Nº1230897 and 1240204, ANID Doctoral Fellowship Grant N° 21232285, ANID Fondecyt Postdoctoral Fellowship Grant N°3260647 and Millennium Science Initiative Program ICN17$_-$012.

\section*{Data availability} 
The code used to generate the figures and support the findings of this article is openly available \cite{salinas.2025.code}.

\section*{Appendix} 
\appendix

\section{Decoherence state deduction}
\label{app:A}

The amplitude damping channel describes the dissipation of energy from a quantum system. This process at zero temperature can be mathematically represented using Kraus operators, denoted as $A_k$. These operators map an initial state $\hat\rho(0)$ to a final state $\hat\rho(\tau)$ after a time $\tau$, according to the following equation:
\begin{equation}
    \hat\rho(\tau)=\sum_{k}A_k\hat\rho(0)A_k^{\dagger}
\end{equation}
where $A_k$ are the Kraus operators defined as:
\begin{equation}
    A_k(\eta)=\sum_{i=k}\sqrt{\binom{i}{k}}\eta^{(i-k)/2}(1-\eta)^{k/2}\ket{i-k}\bra{i}
\end{equation}
Here, $\eta=e^{-\gamma \tau}$ represents the channel's parameter, with $\gamma$ being the damping rate, a measure of how quickly the system loses energy. The index $i$ sums all possible Fock energy levels of the system and $k$ represents the number of photon losses during the damping process.

When considering an initial state $\ket{\alpha_{\varepsilon,t}}$, the action of a Kraus operator $A_k$ can be derived as follows:
\begin{equation}
    \begin{split}
        A_k \ket{\alpha_{\varepsilon,t}}&=\sum_{j=k}^{\infty}\sqrt{\binom{j}{k}}\eta^{(j-k)/2}(1-\eta)^{k/2}\ket{j-k}\bra{j}e^{-it\hat{n}^\varepsilon}\ket{\alpha}\\
        &=e^{-\frac{|\alpha|^2}{2}}\frac{(1-\eta)}{\sqrt{k!}}^{k/2}\sum_{n=0}^{\infty}\eta^{l/2}e^{-it(l+k)^\varepsilon} \frac{\alpha^{l+k}}{\sqrt{l!}}\ket{l}\\
        &= \frac{(\alpha\sqrt{1-\eta})}{\sqrt{k!}}^k e^{-\frac{|\alpha|^2(1-\eta)}{2}} e^{-it(\hat{n}+k)^\varepsilon}\ket{\alpha\sqrt{\eta}}.
    \end{split}
\end{equation}
The application of the Kraus operators to the initial state ultimately leads to an incoherent superposition of GCSs, as detailed in Eq. (\ref{eq:decoherence_GCS}). 
\bibliography{main}

\end{document}